\documentclass[aps,prb,twocolumn,floatfix,footinbib,showpacs]{revtex4-1}
\usepackage{graphicx}
\usepackage{amsfonts,amsmath,amssymb}
\usepackage{amsthm}
\usepackage{dsfont,bm}
\usepackage{color}
\usepackage{soul} %Include when comparing comments
\usepackage{amsbsy}
\usepackage[colorlinks=true,linkcolor=blue,pagecolor=blue,filecolor=blue,menucolor=blue,urlcolor=blue,citecolor=blue,anchorcolor=blue]{hyperref}%
\usepackage{sidecap}
\usepackage{txfonts}
\usepackage{amsbsy}
\usepackage{times} 
\usepackage{dcolumn}

\newcommand{\ua}{\uparrow}
\newcommand{\da}{\downarrow}

\usepackage{amssymb,amsmath,bm}

\begin{document}

\title{Self-biased current, magnetic interference response, and superconducting vortices in tilted Weyl semimetals with disorder}

\author{Mohammad Alidoust}
\affiliation{Department of Physics, K.N. Toosi University of Technology, Tehran 15875-4416, Iran}

\begin{abstract}
We have generalized a quasiclassical model for Weyl semimetals with a tilted band in the presence of an externally applied magnetic field. This model is applicable to ballistic, moderately disordered, and samples containing a high density of nonmagnetic impurities. We employ this formalism and show that a self-biased supercurrent, creating a $\varphi_0$-junction, can flow through a triplet channel in Weyl semimetals. Furthermore, our results demonstrate that multiple supercurrent reversals are accessible through varying junction thickness and parameters characterizing Weyl semimetals. We discuss the influence of different parameters on the Fraunhofer response of charge supercurrent, and how these parameters are capable of shifting the locations of proximity-induced vortices in the triplet channel.   
\end{abstract}

%==========
\date{\today} 

\maketitle
%=========
\section{introduction}
The topological state of matter has been a striking topic during the past decade and attracted extensive attention
both theoretically and experimentally \cite{rev1}. The topological phases can host topologically protected intriguing phenomena and exotic particles, which offer promising prospects to practical arena such as topological 
quantum computation \cite{rev1}. The research efforts in this context have so far been fruitful and led to the exploration of topological insulators \cite{ti1,ti2} and Weyl semimetals \cite{wyl1,wyl2,wyl3,wyl5,wyl6,wyl7,Xu1}, for instance. A topological insulator possesses insulating characteristics in its bulk material and shows perfect conducting features in its surface channels. Also, the band touching points of Weyl semimetals are the so-called Weyl nodes where the Fermi surface, encompassing the nodes, has a nonzero Chern number, thus topologically is nontrivial. The interplay of topological phase with superconductivity is expected to result in topological superconductivity, hosting Majorana fermions governed by non-Abeian statistics\cite{ramon,Shapiro,Banerjee}.  

The conventional BCS superconductivity in Weyl semimetals can occur due to
the inter-valley couplings while the unconventional triplet
correlations may arise by the intra-valley pairings \cite{sc1_theor,sc2_theor,sc3_theor,sc4_theor,sc5_theor,sc6_theor,sc7_theor,sc8_theor,
sc9_theor,sc10_theor,sc12_theor,sc13_theor,alidoustBP1,alidoustBP2,Gorbar,Zhou11}. The
latter case, if energetically favorable, might create superconducting
correlations with finite momentum that places these correlations in the
Fulde-Ferrell-Larkin-Ovchinnikov (FFLO) phase \cite{sc1_theor,sc5_theor}. The
FFLO phenomenon was first predicted for conventional BCS superconductors near their
critical magnetic field where the BCS superconductivity is suppressed
and the amplitude of singlet Cooper pairs is highly
oscillatory\cite{fflo}. Considering the topological phase transition, which
is inherent to the bulk material of ballistic Weyl semimetals, the existence of
superconductivity in such materials can provide a unique platform to
reveal the interplay of superconductivity and topology \cite{sc1_exp,sc2_exp,sc3_exp,vol1,vol2} that has both theory and experiment attention\cite{Sirohi,Soluyanov_app,Ruan_app,Zhang,Zhang2,Chen,Xu,Fang,alidoustBP1,alidoustBP2, Bovenzi,Kononov,Kononov2,Bachmann_app,Xiao_app,shapiro2,Hou_app,Li,Das,wylsuperc_exp1,Teknowijoyo,Lu,Guguchia,Takahashi,QiaoLi}. In particular, quite recent experimental progresses have observed enhancements
in the critical temperature of superconducting $\rm MoTe_2$, from
$0.1$~K to $8.2$~K, under pressures of the order of $11.0$~GPa or from $0.1$~K to
$1.3$~K through partially substituting the tellurium ions by sulfur
\cite{sc1_exp,sc2_exp}. This enhancement is attributed to the
interplay of topology and superconductivity \cite{sc1_exp,sc2_exp}. Nonetheless, this enhancement might be due to the emergence of type-II Weyl semimetal phase in which the transition from type-I to type-II phase increases the available density of states near the Weyl nodes as recently explored in theory \cite{MA_type2,shapiro,volovik}. This transition can be achieved by tensile stress or dopping \cite{MA_type2,shapiro,volovik}. 

Experimentally, the presence of disorder and nonmagnetic impurities in majority of samples is inevitable and may highly influence data analyses of physical quantities sensitive to them. A prominent example is the surface of topological insulators that are expected to be ballistic, showing conductance values equal to those of theory predictions. However, experimental measurement of the conductance of these surface channels was inconsistent with theory predictions. This seemingly discrepancy was resolved through magnetic scanning methods and further conductance spectroscopy analyses. It was demonstrated that disorder and impurities in these surface channels are practically unavoidable and highly alter the conductance of these channels \cite{exp_ti4c}. To properly model realistic surface channels of topological insulators with different densities of nonmagnetic impurities, a quasiclassical approach was recently generalized in the presence of superconductivity and arbitrary magnetization patterns, addressing both equilibrium and non-equilibrium states \cite{zu1,zu2,zu3}. Likewise, the focus of literature has so far been ideal systems and less attention paid to disordered Weyl semimetals in the presence of superconductivity. In this paper, we develop a quasiclassical model for Weyl semimetals with the inclusion of a tilting parameter \cite{MA_type2} and derive Eilenberger and Usadel equations for superconducting Weyl semimetals subject to an externally exerted magnetic field. The Eilenberger equation supports ballistic systems and samples with a moderate density of impurities while the Usadel model covers samples with a high density of impurities and disorder that make the motion of quasiparticles \textit{diffusive}. As a practical application of the developed model, we apply the Usadel equation to Weyl semimetal mediated Josephson junctions. We study crossovers in charge supercurrent and demonstrate that a spontaneous supercurrent can flow through a triplet channel, creating a $\varphi_0$-junction, which is well controlled via junction length and the material parameters pertaining to Weyl semimetals that provide experimentally efficient control knobs. We also consider a two-dimensional junction subject to an external magnetic field and show that the charge current has a decaying oscillatory behavior by increasing the external magnetic field, constituting Fraunhofer-modulated diffraction patterns. In all cases, we evaluate the influence of the tilting parameter on our findings. 

The paper is organized as follows. In Sec. \ref{sec:1}, starting from the low energy Hamiltonian, we present main steps of formulating a quasiclassical model for Weyl semimetals. Considering a standard model for nonmagnetic impurities, we derive the Eilenberger equation which is applicable to ballistic and moderately disordered samples. Next, we consider samples with a high density of impurities, average the Eilenberger equation over particles' momentum, and derive the Usadel equation. We then discuss the tunneling boundary conditions and derive a charge current relationship for the model Hamiltonian we start with. In Sec. \ref{sec:1D}, we consider a Josephson configuration, find solutions to the Green's function, and derive an analytical expression to the supercurrent phase relation. We study the spontaneous supercurrent, current reversals, and the effects of tilting parameter on them. In Sec. \ref{sec:2D}, we consider a two-dimensional junction subject to an external magnetic field and study the behavior of supercurrent flow and superconducting vortices. Finally, we give concluding remarks in Sec. \ref{sec:cncl}.     

\section{Eilenberger and Usadel equations }\label{sec:1}

We first discuss the Hamiltonian of normal state Weyl semimetal and next incorporate superconductivity. The model Hamiltonian that governs the dynamics of low energy particles inside a ballistic Weyl semimetal subject to an external magnetic field reads:
\begin{equation}\label{hamil}
\begin{split}
&H=\sum_{\sigma\sigma'}\int \frac{d\textbf{k}}{(2\pi)^3} \psi_\sigma^\dag(\textbf{k}) \Big\{ \gamma \Big[(\text{k}_k+e\text{A}_k)^2-Q^2\Big]\sigma_z\\&+\beta \Big[(\text{k}_k+e\text{A}_k)^2-Q^2\Big]+\alpha_{k}{(\text{k}_k+e\text{A}_k)}{ \sigma_k} -\mu\Big\}_{\sigma\sigma'} \psi_{\sigma'}(\textbf{k}),~~~~~~~
\end{split}
\end{equation}
in which indices $\sigma,\sigma'\equiv\uparrow,\downarrow$, $k\equiv x,y,z$, $\gamma$ characterizes Weyl semimetal and breaks the time reversal symmetry, $Q$ is the splitting of Weyl nodes, $\beta$ describes the tilt of the Weyl cones, $\alpha_{k}$ is the strength of the inversion symmetry breaking parameter, and $\mu$ stands for the chemical potential. The particles' momentum is denoted by $\textbf{k}=(\text{k}_x,\text{k}_y,\text{k}_z)$ and the external magnetic field is given through its associated vector potential $\textbf{A}=(\text{A}_x,\text{A}_y,\text{A}_z)$. Here, ${\bm \sigma}=(\sigma_x,\sigma_y,\sigma_z)$ are the Pauli matrices and $e$ is the elementary charge.

To describe a system made of Weyl semimetal, we define propagators;
\begin{subequations}\label{GF_comps}
\begin{eqnarray}
G_{\sigma\sigma'}(t,t'; \mathbf{ r}, \mathbf{ r}') &=& -\langle {\cal T}\Psi_{\sigma} (t,\mathbf{ r}') \Psi_{\sigma'}^{\dag}(t',\mathbf{ r}')  \rangle,~~~~~~~\\
\bar{G}_{\sigma\sigma'}(t,t'; \mathbf{ r}, \mathbf{ r}') &=& - \langle {\cal T} \Psi^{\dag}_{\sigma} (t,\mathbf{ r}) \Psi_{\sigma'}(t',\mathbf{ r}')  \rangle,~~~~~~~\\
F_{\sigma\sigma'}(t,t'; \mathbf{ r}, \mathbf{ r}') &=& + \langle {\cal T} \Psi_{\sigma} (t,\mathbf{ r}) \Psi_{\sigma'}(t',\mathbf{ r}')  \rangle,~~~~~~~\\
F^{\dag}_{\sigma\sigma'}(t,t'; \mathbf{ r}, \mathbf{ r}') &=& + \langle {\cal T} \Psi^{\dag}_{\sigma} (t,\mathbf{ r}) \Psi^{\dag}_{\sigma'}(t',\mathbf{ r}')  \rangle,~~~~~~~
\end{eqnarray}
\end{subequations}
where ${\cal T}$ is the time ordering operator, $t, t'$ are the imaginary times at $\mathbf{ r}, \mathbf{ r}'$, respectively.
We consider elastic impurity scattering potentials $V(\mathbf{ r})$ inside Weyl semimetal by a self-energy term
\begin{equation}
\Sigma_{\text{imp}}(\mathbf{ r},\mathbf{ r}')=\langle V(\mathbf{ r})G(\mathbf{ r},\mathbf{ r}')V(\mathbf{ r}')\rangle,
\end{equation}
where we average over the positions of impurities. To obtain the above self-energy term, we treat the impurity potentials as perturbation and expand the Green's function in terms of the unperturbed Green's function up to the second order. We find the mean free time of particles in the disordered Weyl semimetal as $\tau^{-1}=2\pi n_i N_0\int d\Omega_{{\bm n}_\text{F}}(4\pi)^{-1}|v(\Omega)|^2$ in which $v(\Omega)$ is the Fourier transform of the scattering potential that depends on the relative angle $\Omega$ between the incident and scattered direction of particles, $N_0$ is the density of states per spin at the Fermi level of the system and $n_i$ is the concentration of impurities. Note that, for the sake of simplicity in the subsequent calculations, we have neglected the intervalley scatterings and anisotropic terms and their effects on the mean free time. In the particle-hole space we find the following equation for the Green's function:
\begin{eqnarray}\nonumber \label{Green1}
\left( \begin{array}{cc}
-i\omega_n+\hat{H}(\mathbf{ r})& -\hat{\Delta}(\textbf{r}) \\
\hat{\Delta}^*(\textbf{r}) & i\omega_n +\sigma_y\hat{H}^*(\mathbf{ r})\sigma_y
\end{array} \right)\check{G}(\omega_n;\mathbf{ r},\mathbf{ r}')
\\
=\delta(\mathbf{ r}-\mathbf{ r}')+\frac{1}{2\pi N_0 \tau} \check{G}(\omega_n;\mathbf{ r},\mathbf{ r}) \check{G}(\omega_n;\mathbf{ r},\mathbf{ r}'),
\end{eqnarray}
in which $\omega_n=\pi (2n+1)k_BT$ is the Matsubara frequency, $n\in { Z}$, $k_B$ is the Boltzman constant, $T$ is temperature, and $\hat{\Delta}(\mathbf{r})$ is the superconducting gap inside Weyl semimetal. The matrix form of the Green's function is given by;
\begin{equation}
\nonumber \check{G}(\omega_n;\mathbf{ r},\mathbf{ r}')=\left(  \begin{array}{cc}
-\hat{G}(\omega_n;\mathbf{ r},\mathbf{ r}') & -i\hat{F}(\omega_n;\mathbf{ r},\mathbf{ r}')\sigma_y\\
-i\sigma_y\hat{F}^\dag(\omega_n;\mathbf{ r},\mathbf{ r}') & \sigma_y\hat{\bar{G}}(\omega_n;\mathbf{ r},\mathbf{ r}')\sigma_y
\end{array}  \right).
\end{equation}
We have denoted $2\times 2$ matrices by `hat' symbol, $\hat{\square}$, and $4\times 4$ matrices by `check' symbol, $\check{\square}$.  
Next, we subtract from Eq. (\ref{Green1}) its conjugate and perform a Fourier transformation with respect to the relative coordinates: $\mathbf{ R}= (\mathbf{ r}+\mathbf{ r}')/2$ and $\delta \mathbf{ r} = \mathbf{ r}-\mathbf{ r}'$. In order to simplify our calculations, we assume that the Fermi energy is the largest energy scale in the system, and thus the Green's function is localized at the Fermi level (with Fermi velocity $v_F$). Hence, define the quasiclassical Green's function 
\begin{eqnarray}\label{G2}
\check{g}(\omega_n; \mathbf{ R}, \mathbf{ n}_\text{F}) =\frac{i}{\pi}\int d\xi_\text{p} \check{G}(\omega_n; \mathbf{ R}, \mathbf{ p}),
\end{eqnarray}
in which $d\xi_\text{p}=v_Fd\text{p}$. Incorporating these assumptions we finally arrive at the Eilenberger equation \cite{eiln}:
%\begin{equation}
\begin{gather}
\text{p}_\text{F}^k\Big\{{\cal L}, {\check{\tilde{\nabla}}_k \check{g}}\Big\}
 +\Big[\omega_n\tau_z+\check{\Gamma}_k+\frac{1}{2\tau} \langle \check{g} \rangle,\check{g}\Big]=0,\nonumber\\
 {\cal L}=\gamma\sigma_z+\beta,\label{eilenb}
 \\ \nonumber\check{\tilde{\nabla}}_k {\check{\text{X}}}\equiv \check{\nabla}_k {\check{\text{X}}} - \Big[ie\text{A}_k\tau_z,{\check{\text{X}}}\Big],\\\nonumber
 \check{\Gamma}_k=-i\check{\Delta}-i\text{p}_\text{F}^k{\alpha_{k}} \tau_z{ \sigma}_k+i\gamma Q^2\sigma_z+i\beta Q^2,
\end{gather}
where ${\bm \tau}=(\tau_x,\tau_y,\tau_z)$ are Pauli matrices in particle-hole space, $\mathbf{p}_\text{F}=(\text{p}_\text{F}^x,\text{p}_\text{F}^y,\text{p}_\text{F}^z)$, and $\check{\nabla}_k\equiv \check{\partial}_{x,y,z}$. The average over disorder is shown by $\langle ... \rangle$. In the above calculations, we have neglected contributions of the order of $L^{-2}$ and $\omega_c/\mu \ll 1$, where $\omega_c$ is the cyclotron frequency and $L$ is a length scale, large enough compared to the Fermi wave length; $L\gg \lambda_\text{F}$. Also, it is assumed that ${\cal L}$ is the leading term in the Hamiltonian.   

\begin{figure}[t!]
\includegraphics[clip, trim=4cm 2.8cm 4cm 4cm, width=8.0cm,height=6.0cm]{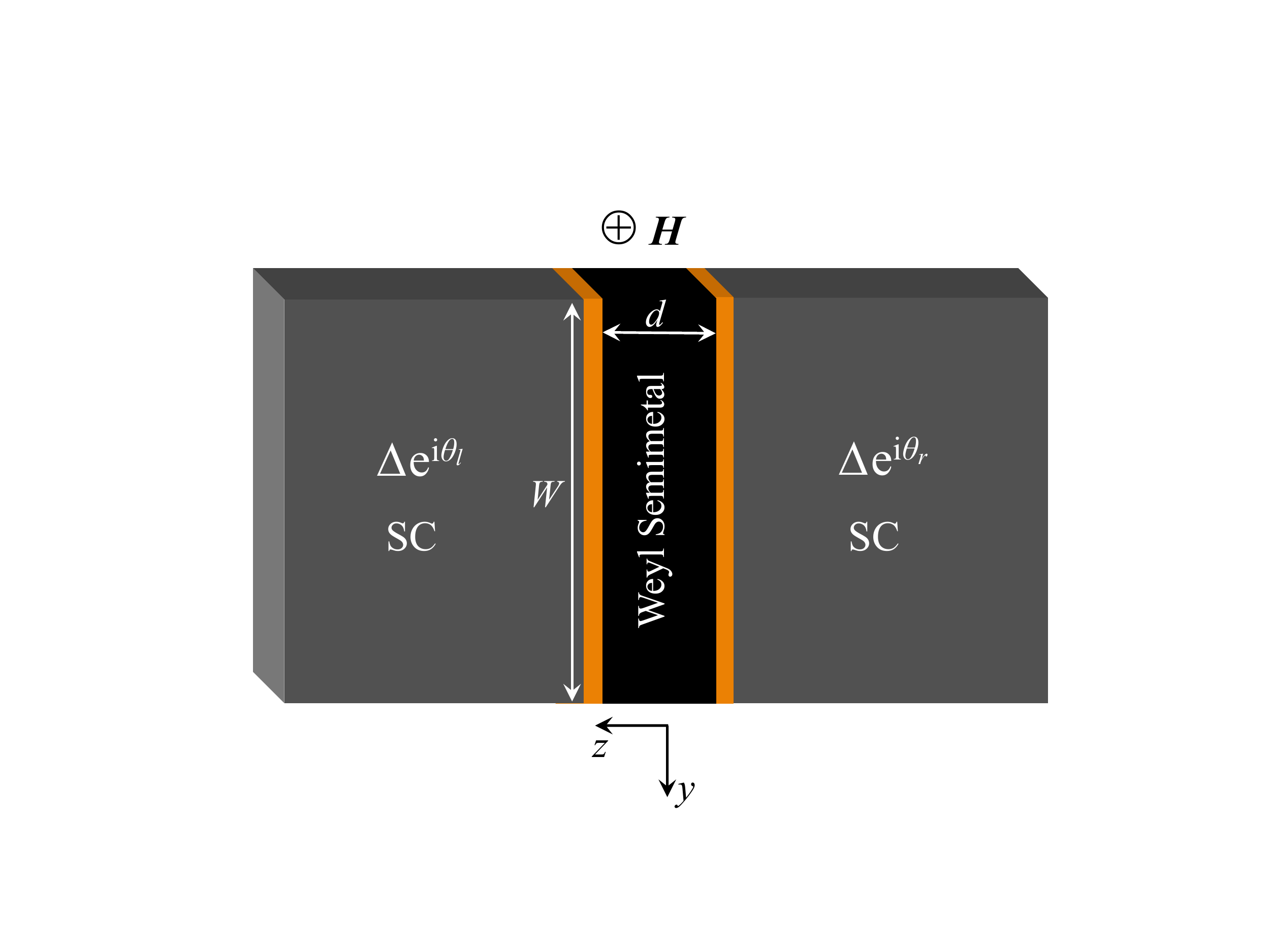}
\caption{\label{fig1} (Color online). 
Schematic of a Josephson junction made of a disordered Weyl semimetal. The junction plane is placed in the $zy$ plane and the tunneling interfaces between the Weyl semimetal and superconductors are located at $z=\pm d/2$. The macroscopic phase of the left and right superconductors are labeled by $\theta_l$ and $\theta_r$, respectively. The width of junction is $W$ and an external magnetic field ${\bm H}$ is applied along the $x$ axis, perpendicular to the junction plane.}
\end{figure}

The Eilenberger equation can be more simplified in systems with numerous impurities so that $ 1/\tau > |\omega_n| , |\Delta|$. In this case, the quasiparticles move diffusively with random directions and trajectories, which is the so-called diffusive regime \cite{usadel}. 
In the diffusive regime, we integrate the quasiclassical Green's function, Eq. (\ref{G2}), over all possible directions of quasiparticles' momentum:
\begin{eqnarray}
\langle \check{g}(\omega_n; \mathbf{ R}, \mathbf{ n}_\text{F}) \rangle \equiv \int\frac{d\Omega_{\mathbf{ n}_\text{F}}}{4\pi}\check{g}(\omega_n; \mathbf{ R}, \mathbf{ n}_\text{F}), \;\mathbf{ n}_\text{F} =\frac{\mathbf{ p}_\text{F}}{|\mathbf{ p}_\text{F}|}. ~~~~~\;\;
\end{eqnarray}
In this regime, the Green's function can be expanded through the first two harmonics: $s$-wave and $p$-wave
\begin{equation}\label{expansion}
\check{g} (\omega_n; \mathbf{ R},\mathbf{ n}_\text{F}) = \check{\text{g}}_s(\omega_n; \mathbf{ R}) + { n}_\text{F}^k \check{\text{g}}_p^k (\omega_n; \mathbf{ R}),
\end{equation}
where the $s$-wave harmonic in the expansion above (\ref{expansion}) is isotropic and its magnitude is much larger than the $p$-wave harmonic: $\check{\text{g}}_s \gg  \check{\text{g}}_p^k$. By substituting this expanded Green's function into Eq. (\ref{eilenb}) and performing an integration over momentum directions we find
\begin{eqnarray}\label{Int1}
&\check{\text{g}}_p^k=-\tau\text{p}_\text{F}^k\check{\text{g}}_s\Big\{ {\cal L},  \check{\tilde{ \nabla}}_k \check{\text{g}}_s \Big\}+\tau\text{p}_\text{F}^k\check{\text{g}}_s\Big[i\alpha_{k} \tau_z{ \sigma}_k , \check{\text{g}}_s \Big].~~~~
\end{eqnarray}
Next, by substituting Eq. (\ref{Int1}) into Eq. (\ref{eilenb}) and assuming that ${\nabla}_k\gamma={\nabla}_k\beta={\nabla}_k\alpha_{k}={\nabla}_k\text{A}_k=0$, we find a generalized Usadel model for tilted Weyl semimetals \cite{usadel}: 
\begin{equation}\label{Usadel}
\frac{\text{p}_\text{F}^k}{3}\Big\{ {\cal L}, \check{\tilde{ \nabla}} _k \check{\text{g}}_p^k\Big\}-\frac{\text{p}_\text{F}^k}{3}\Big[  i\alpha_{k} \tau_z{ \sigma}_k, \check{\text{g}}_p^k\Big]+\Big[\omega_n \tau_z+\check{\Gamma}_k,\check{\text{g}}_s \Big]=0.~~~
\end{equation}

Solving Eqs. (\ref{eilenb}) and (\ref{Usadel}), one finds appropriate Green's function that contains all information describing observable physical properties of various systems. Hence, we now proceed to apply our formulated quasiclassical model to hybrid structures made of disordered Weyl semimetals and superconductors, which are practical platforms that the Eilenberger and Usadel equations are able to describe them properly. We note that the quasiclassical Eilenbeger and Usadel approaches were also generalized for spin-orbit coupled systems \cite{Konschelle,Bergeret,Huang,ali_so1,ali_so2,gupta}, surface channels of topological insulators \cite{zu1,zu2,zu3}, and black phosphorus monolayer (phosphorene) \cite{alidoustBP2} in the presence of superconductivity and a Zeeman field. A specific configuration is depicted in Fig. \ref{fig1}. As seen, two superconductors are coupled through a disordered Weyl semimetal of thickness $d$ and width $W$. The interfaces are located at $z=\pm d/2$ in the $xy$ plane. The macroscopic phases of the left and right superconductor terminals are denoted by $\theta_l$ and $\theta_r$, respectively.  
We consider low transparent interfaces (the so-called tunneling limit) between the superconductors and Weyl semimetal and neglect the inverse proximity effect at the interfaces. We therefore find the following expression to the boundary conditions \cite{zu1,zu2,zu3,boundary_c1,ma_jap}
\begin{equation}\label{BC_sup}
\zeta { n}_k\check{\text{g}}_p^k = \Big[\check{\text{g}}_s, \check{\text{g}}_{\text{SC}}\Big],
\end{equation}
in which ${ n}_k $ is a unit vector perpendicular to a boundary, $\zeta$ controls the opacity of interfaces, and $\check{\text{g}}_{\text{SC}}$ is the Green's function of the bulk superconductors. To study the quantum transport, we derive an expression to the charge current flow (due to the superconducting phase gradient across the device, in our case). The quantum definition of current density is expressed through the Hamiltonian Eq. (\ref{hamil}) as follows
\begin{equation}\label{crnt_hamil}
\begin{split}
\frac{\partial \rho}{\partial t}=\lim\limits_{{\bm r}\rightarrow {\bm r}'}\sum\limits_{\sigma\sigma'}\frac{1}{i\hbar}\Big[ \psi^\dag_{\sigma}({\bm r}')H_{\sigma\sigma'}({\bm r})\psi_{\sigma'}({\bm r})\\-\psi^\dag_{\sigma}({\bm r}')H_{\sigma\sigma'}^\dag({\bm r}')\psi_{\sigma'}({\bm r})\Big],
\end{split}
\end{equation}
where the left hand side is the time variation of charge density $\rho$. Throughout our calculations, we consider a steady state regime and, therefore, set $\partial \rho/\partial t=0$. We use the Fourier representation of the Keldysh Green’s function in equilibrium:
\begin{equation}\label{greenKt}
\sum_n\int\frac{d\textbf{p}}{(2\pi)^3}e^{i\textbf{p}\cdot\textbf{r}}\check{G}^K(\omega_n; \textbf{R},\textbf{p}),
\end{equation}
and we finally arrive at an expression for the current density in the ballistic regime. By applying the quasiclassical approximations described above and making use of the harmonics expansion to the Green's function, Eq. (\ref{expansion}), we find the following expression for the current density in the diffusive regime: 
\begin{equation}\label{crnt_2}
{J}_k = \frac{ie\pi }{3}N_0\text{p}_\text{F}^kT \sum_n \mathrm{Tr} \Big[\tau_z \big(\gamma\sigma_z+\beta\big)\check{\text{g}}_p^k\Big].
\end{equation}
To derive Eq. (\ref{crnt_2}) we have assumed sufficiently small $\alpha_{k}$ and neglected terms of the order of $\alpha_{k}(\text{p}_\text{F}^k)^{-1}$.

\section{self-biased supercurrent and supercurrent reversals}\label{sec:1D}
In order to find the charge current density, one has to solve either Eq.(\ref{eilenb}) (in the ballistic regime\cite{zu3}) or Eq. (\ref{Usadel}) (in the diffusive regime\cite{zu1,zu2}) together with proper boundary conditions (\ref{BC_sup}) and substitute the resultant Green's function into Eq. (\ref{crnt_hamil}). In the diffusive regime, the Usadel equation (\ref{Usadel}) results in nonliner boundary value differential equations that has to be solved numerically \cite{ma_odfr}. To obtain decopled linear differential equations that are simpler to solve and provide analytical solutions, we expand and linearize the Green's function aournd the bulk solution $\check{\text{g}}_0(\omega_n;\mathbf{R})$, i.e., $\check{\text{g}}(\omega_n;\mathbf{R})\approx \check{\text{g}}_0(\omega_n;\mathbf{R})+\check{f}(\omega_n;\mathbf{R})$\cite{ma_jap}. This approximation is experimentally relevant and accessible in a low proximity limit either close to the superconducting critical temperature or devices with low transparent interfaces \cite{ma_jap}. The external magnetic field is given by its associated vector potential that satisfies the Lorentz gauge ${\bm \nabla}\cdot\textbf{A}=0$ and ${H}_x={\bm \nabla}\times\textbf{A}$. As depicted in Fig. \ref{fig1}, we consider a situation where the external magnetic field is directed towards $x$ direction, perpendicular to the junction plane, and, therefore, can be described by $\textbf{A}=(0,0,y{ H}_x)$. The Usadel equation (\ref{Usadel}) for the triplet channel in the presence of the external magnetic field within a tilted Weyl semimetal results in the following decoupled linear differential equations: 
\onecolumngrid
%\begin{widetext}
\begin{subequations}\label{usadl2d}
%\small
\begin{eqnarray}
 & (\beta + \gamma)^2{ \nabla}^2_kf_{\ua\ua}(\omega_n)-[\alpha_{z}+2e\text{A}_z(\beta+\gamma)]^2 f_{\ua\ua}(\omega_n)  - 2 i (\beta + \gamma)[\alpha_{z}+2e\text{A}_z(\beta+\gamma)]  { \nabla}_z f_{\ua\ua} (\omega_n)
  +  \omega_nD^{-2} f_{\ua\ua}(\omega_n)=0,~~~~~~~~\\
 & (\beta - \gamma)^2{ \nabla}^2_kf_{\da\da}(\omega_n)- [\alpha_{z}-2e\text{A}_z(\beta-\gamma)]^2 f_{\da\da}(\omega_n)  +2 i ( \beta - \gamma) [\alpha_{z}-2e\text{A}_z(\beta-\gamma)] { \nabla}_z f_{\da\da} (\omega_n)
   +  \omega_nD^{-2} f_{\da\da}(\omega_n)=0,~~~~~~~~\\
 & (\beta + \gamma)^2 { \nabla}^2_k\tilde{f}_{\ua\ua}(\omega_n) -[\alpha_{z}+2e\text{A}_z(\beta+\gamma)]^2 \tilde{f}_{\ua\ua}(\omega_n)   +2 i  (\beta + \gamma)[\alpha_{z}+2e\text{A}_z(\beta+\gamma)] { \nabla}_z \tilde{f}_{\ua\ua}(\omega_n) 
  +  \omega_nD^{-2} \tilde{f}_{\ua\ua}(\omega_n)=0,~~~~~~~~\\
 &  (\beta - \gamma)^2 { \nabla}^2_k\tilde{f}_{\da\da}(\omega_n)-[\alpha_{z}-2e\text{A}_z(\beta-\gamma)]^2 \tilde{f}_{\da\da}(\omega_n)  - 2 i (\beta - \gamma) [\alpha_{z}+2e\text{A}_z(\beta-\gamma)] { \nabla}_z \tilde{f}_{\da\da}(\omega_n) 
   +  \omega_nD^{-2} \tilde{f}_{\da\da}(\omega_n)=0.~~~~~~~~
\end{eqnarray}
%\normalsize
\end{subequations}
%\end{widetext}
\twocolumngrid
Here, index $k$ runs over $y,z$ in a two-dimensional system and $x,y,z$ in a three-dimensional junction.

\begin{figure*}[t!]
\includegraphics[width=17.50cm,height=7.0cm]{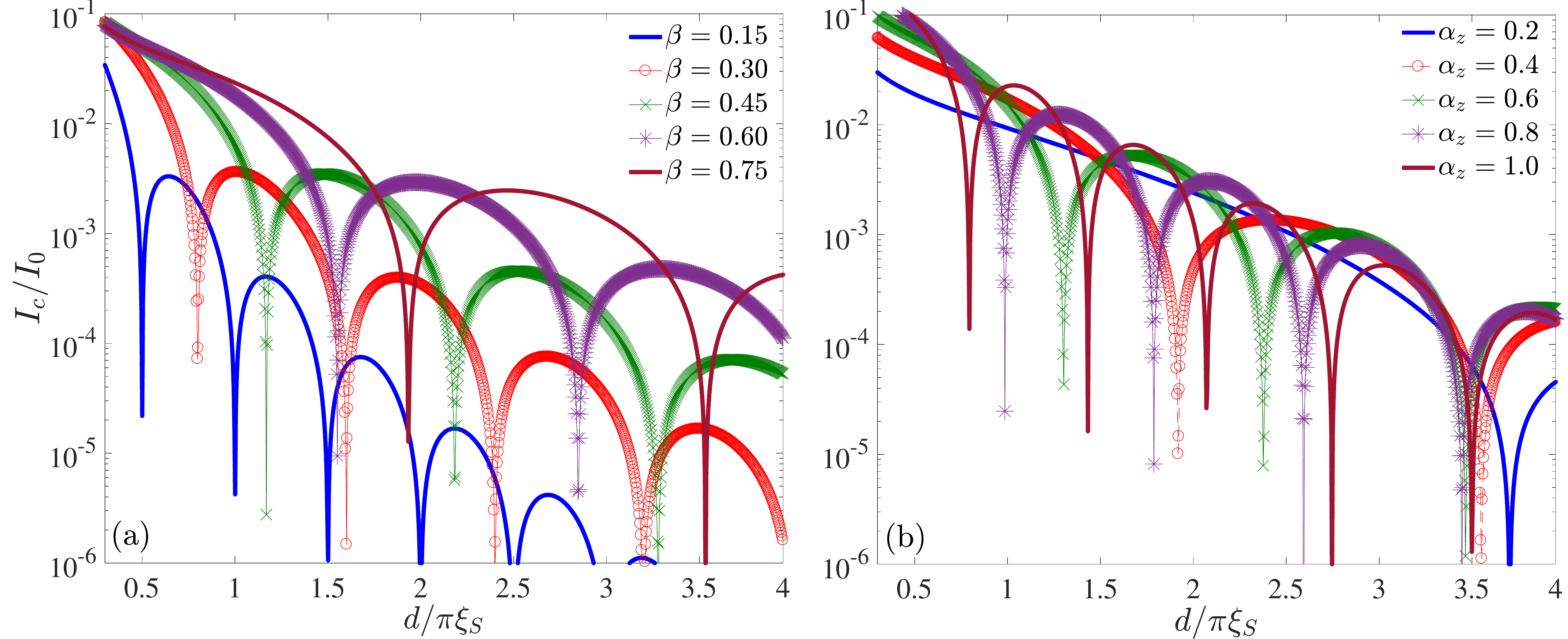}
\caption{\label{fig2} (Color online). 
Normalized charge current through the disordered Weyl semimetal Josephson junction as a function of junction thickness $d(\pi\xi_S)^{-1}$. (a) The supercurrent is shown for differing values of $\beta=0.15, 0.30, 0.45, 0.60$, and $0.75$ with fixed $\gamma=0.1$ and $\alpha_{z}=0.5$ fixed. (b) The supercurrent is shown for differing values of $\alpha_{z}=0.2, 0.4, 0.6, 0.8$, and $1.0$ with fixed $\gamma=0.1$ and $\beta=0.6$. 
 }
\end{figure*}

To begin, we first set the external magnetic field zero, i.e., ${ H}_x=0$ in Eqs. (\ref{usadl2d}). By considering the low proximity limit, described above, we find appropriate expressions to the components of Green's function. For example, the components $f_{\ua\ua}(\omega_n; z)$ and $\tilde{f}_{\ua\ua}(\omega_n; z)$ become:
\onecolumngrid
%\begin{widetext}
\begin{subequations}
%\begin{equation}
\begin{align}
&&\begin{split}
&f_{\ua\ua}(\omega_n; z)={\cal F}(\omega_n) \exp \left( {-\frac{2 i \alpha_z d z}{\beta +\gamma}}\right) \left\{\exp \left(\frac{i (\alpha_z
   d (z+1)+\theta_r (\beta +\gamma)+\lambda_nD^{-1} (1-z))}{\beta
   +\gamma}\right)+\right.\\&\left.\exp \left(\frac{i (\alpha_z d (z+1)+\theta_r (\beta
   +\gamma)+\lambda_nD^{-1} (z+1))}{\beta +\gamma}\right)+ \exp \left( {\frac{i (\alpha_z d
   z+\beta  \theta_l+\gamma \theta_l-\lambda_nD^{-1} (z-2))}{\beta
   +\gamma}}\right)+\exp \left(  i\frac{z (\alpha_z d+\lambda_nD^{-1})}{\beta
   +\gamma}+i\theta_l\right)\right\},
\end{split}\\
&&\begin{split}
&\tilde{f}_{\ua\ua}(\omega_n; z)={\cal F}(\omega_n) \exp\left(-i\frac{\alpha_z d+(\beta +\gamma)(\theta_l+\theta_r)}{\beta +\gamma}\right)
   \left\{\exp \left(\frac{i (\alpha_z d (z+1)+\theta_r (\beta
   +\gamma)-\lambda_nD^{-1} (z-2))}{\beta +\gamma}\right)+\right.\\&\left. \exp \left({\frac{i (\alpha_z d
   z+(\beta+\gamma)  \theta_l+\lambda_nD^{-1} (1-z))}{\beta
   +\gamma}}\right)+ \exp \left({\frac{i (\alpha_z d z+(\beta+\gamma)  \theta_l+\lambda_nD^{-1} (1+z))}{\beta +\gamma}}\right)+\exp\left({\frac{i (\alpha_z d
   (z+1)+(\beta+\gamma)  \theta_r+\lambda_nD^{-1} z)}{\beta
   +\gamma}}\right)\right\} . 
\end{split}
\end{align}
\end{subequations}
%\end{widetext}
\twocolumngrid
Here, we have defined ${\cal F}(\omega_n)=iD\Big\{\lambda_n\zeta[\exp \left({\frac{2i\lambda_nD^{-1}}{\beta+\gamma }}\right)-1]\Big\}^{-1}f_t$, $D^2=2\text{p}_\text{F}^2\tau/3$, and $\lambda_n^2=-\omega_n/d^2$. Our analyses of the boundary conditions and the Usadel equation demonstrate that the supercurrent in this system can flow through a triplet channel. Therefore, we assume a triplet component $f_t$ to $\hat{\text{g}}_{\text{SC}}$ in Eq. (\ref{BC_sup}). This finding may explain a recent experiment where a long-ranged supercurrent was observed through a Josephson junction made of $\rm WTe_2$ Weyl semimetal \cite{Kononov2} inline with previous works \cite{Konschelle,Bergeret,ali_so1,ali_so2,gupta}. In the singlet channel we recover the results of a conventional SNS junction (up to the zero order of $\alpha_z\text{p}_\text{F}^{-1}$). We only presented two representative components of $\tilde{\hat{f}}(\omega_n; z)$ and $\hat{f}(\omega_n; z)$. To obtain the total charge current passing through the junction, $I_c$, we substitute these solutions into Eq. (\ref{crnt_2}) and, after performing calculations, find the following charge supercurrent phase relation  
\onecolumngrid
%\begin{widetext}
\begin{equation}\label{crnt1d}
I_c=e\pi N_0{\cal A} D^3T\sum_n \frac{ f_t^2}{\zeta^2\lambda_n} \csc\Big(\frac{\lambda_nD^{-1}}{\beta-\gamma}\Big) \csc\Big(\frac{\lambda_nD^{-1}}{\beta+\gamma}  \Big)\left\{ (\beta-\gamma)\sin\Big(\frac{\lambda_nD^{-1}}{\beta+\gamma}  \Big)\sin \Big(\frac{\alpha_z d}{\beta-\gamma}-\varphi\Big) - (\beta+\gamma)\sin\Big(\frac{\lambda_nD^{-1}}{\beta-\gamma}  \Big)\sin \Big(\frac{\alpha_z d}{\beta+\gamma}+\varphi\Big)\right\},
\end{equation}
%\end{widetext}
\twocolumngrid
in which ${\cal A}$ is the cross section of the Weyl semimetal/superconductor interface, $\varphi=\theta_l-\theta_r$, and we define $I_0=e\pi N_0{\cal A}$. As seen in Eq. (\ref{crnt1d}), the supercurrent experiences a total phase shift $\Theta_0(\beta,\gamma,\alpha_z)$ made of $\varphi_0^\pm=d\alpha_{z}/(\beta\pm \gamma)$ that renders the junction grand state into values other than the standard $0$ or $\pi$ states in conventional Josephson junctions. This phase shift causes a self-biased supercurrent at zero phase difference $\varphi=0$. Note that $\varphi_0^\pm$ are independent of $D$. This finding illustrates that the addressed phase shift is robust against the density of impurities considered in the system, i.e., $\varphi_0^\pm$ are independent of $\tau$. Hence, this phenomenon can obviously occur in moderately disordered and ballistic regimes as quite recently explored in experiment\cite{herve} inline with theory predictions for topological insulator surface channels\cite{zu1,zu2,zu3} and black phosphorus monolayer \cite{alidoustBP1,alidoustBP2}. Also, the explored $\varphi_0$ state in this paper relies on the inherent parameters of Weyl semimetal without involving Zeeman field \cite{zu1,zu2,zu3,alidoustBP1,alidoustBP2,phi0,herve}. It is worth mentioning that the appearance of $4\pi$-periodic current phase relation in topological insulator Josephson junctions might be due to the presence of Majorana Fermions although such a $4\pi$-periodic supercurrent phase relation can be theoretically obtained in the trivial regime of a ballistic topological insulator \cite{zu3,4pi}. Figure \ref{fig2} illustrates the normalized charge supercurrent as a function of the thickness of Weyl semimetal $d$ normalized by the superconducting coherence length $\xi_S$ for differing values of the tilting parameter $\beta$, Fig. \ref{fig2}(a), and strength of inversion symmetry breaking parameter $\alpha_z$, Fig. \ref{fig2}(b), at zero phase difference, i.e., $\varphi=0$. In Fig. \ref{fig2}(a), considering representative values, we set $\alpha_z=0.5, \gamma=0.1$ fixed and vary $\beta$, whereas in Fig. \ref{fig2}(b), $\beta=0.6, \gamma=0.1$ are set fixed and $\alpha_z$ varies. We see that the supercurrent decays and experiences multiple sign changes as a function of $d$ in both cases. The sign change occurs faster by decreasing $\beta$ and increasing $\alpha_z$. This can be understood by Eq. (\ref{crnt1d}). Increasing $\alpha_z$ or decreasing $\beta$, the phase shift increases and, therefore, by varying $d$, faster oscillations occur. Equation (\ref{crnt1d}) demonstrates that $\beta$, in the absence of $\alpha_{z}$, is unable to induce phase shift in this channel while in the presence of a finite $\alpha_{z}$, the tilting parameter changes the magnitude and sign of the phase shifts, making a total phase shift $\Theta_0(\beta,\gamma,\alpha_z)$. 
The inversion symmetry breaking parameter, $\alpha_z$, may respond to a mechanically exerted deformation efficiently \cite{alidoustBP1,alidoustBP2} and, thus, the spontaneous supercurrent and current reversals might be effectively controllable through external knobs regardless of the density of impurities and disorder present in the system.

The tilting parameter competes with the inversion symmetry breaking parameter in inducing supercurrent reversals. The increase of $\beta$ results in shifting the locations of current reversal points and, in general, reduces the number of crossovers that can appear in a specific interval of junction thickness, Fig. \ref{fig2}(a). This is opposite to the effect of $\alpha_{z}$ shown in Fig. \ref{fig2}(b). Furthermore, by increasing $\beta$, the supercurrent vs the junction thickness enhances and decays slower, whereas $\alpha_{z}$ is unable to influence the degree of supercurrent decay. From Eq. (\ref{crnt1d}), the supercurrent is proportional to $(\beta\pm\gamma)\text{csch}(1/d(\beta-\gamma))\text{csch}(1/d(\beta+\gamma))\text{sinh}(1/d(\beta\mp\gamma))\sin(\varphi_0^\pm)$. It is apparent that $\alpha_z$ is unable to effectively enhance or suppress the magnitude of total supercurrent while $\beta$ and $\gamma$ can highly alter the total charge current. If we set $\alpha_z=0$, the supercurrent as a function of junction thickness $d$ decays with no current reversal, showing long-ranged characteristics\cite{Kononov2}. We now proceed to study Weyl semimetal Josephson junctions subject to an external magnetic field.

\section{Faunhofer response and proximity vortices}\label{sec:2D}

In order to study the response of charge current to an external magnetic field in a Weyl semimetal mediated Josephson junction, we consider a two-dimensional configuration with ${\bm H}=({H}_x,0,0)$ depicted in Fig. \ref{fig1}. We assume that $W\gg d$, define ${ \Phi}=\pi Wd{ H}_x$, ${ \Phi}_0=h/2e$ a quantum flux, and ${\bm \Phi}={\Phi}/{\Phi_0}$ \cite{ma_odfr,ma_jap,zu1}. The resultant Green's function components; $f_{\ua\ua}(\omega_n;z,y), \tilde{f}_{\ua\ua}(\omega_n;z,y)$ and calculated current density flowing in the $z$ direction are given by;

\onecolumngrid
%\begin{widetext}
\begin{subequations}
%\begin{equation}
\begin{align}
&&\begin{split}
&f_{\ua\ua}(\omega_n; z)={\cal F}(\omega_n)\exp \left(-\frac{i (z (\alpha_z
   d+\lambda_nD^{-1})-2 \beta  {\bm \Phi} y
   (z-1)-2 \gamma {\bm \Phi} y
   (z-1))}{\beta
   +\gamma}\right)\left\{\exp \left( {\frac{i (\alpha_z
   d+\theta_r (\beta
   +\gamma)+ \lambda_nD^{-1}(2
   z+1))}{\beta
   +\gamma}}\right)+\right.\\
   &\left. \exp \left(\frac{i
   (\alpha_z d+\theta_r (\beta
   +\gamma)+\lambda_nD^{-1})}{\beta
   +\gamma}\right)+\exp \left(\frac{i
   (\beta  (2
   {\bm \Phi} y+\theta_l)+\gamma (2 {\bm \Phi} y+\theta_l)+2
   \lambda_nD^{-1} z)}{\beta
   +\gamma}\right)+\exp {
   \left(\frac{2i \lambda_nD^{-1}}{\beta
   +\gamma}+2i
   {\bm \Phi} y+i\theta_l\right)}\right\},
\end{split}\\
&&\begin{split}
&\tilde{f}_{\ua\ua}(\omega_n; z)={\cal F}(\omega_n)  \exp \left(-\frac{i (\alpha_z
   (d-d z)+\beta  (2 {\bm \Phi} y
   z+\theta_l+\theta_r)+2
   \gamma {\bm \Phi} y
   z+\gamma
   (\theta_l+
   \theta_r)+\lambda_nD^{-1} z)}{\beta
   +\gamma}\right)
   \left\{\exp \left( {\frac{i (\alpha_z
   d+(\beta +\gamma)
   \theta_r+2 \lambda_nD^{-1} z)}{\beta
   +\gamma}}\right)+\right. \\ & \left. \exp\left({\frac{i
   (\alpha_z d+\theta_r (\beta
   +\gamma)+2 \lambda_nD^{-1})}{\beta
   +\gamma}}\right)+\exp \left(\frac{i
   ((\beta +\gamma) (2
   {\bm \Phi} y+\theta_l)+
   \lambda_nD^{-1} (2z+1))}{\beta
   +\gamma}\right)+\exp\left({\frac{i
   ((\beta +\gamma) (2
   {\bm \Phi} y+\theta_l)+\lambda_nD^{-1})}{\beta +\gamma}}\right)\right\},
\end{split}
\end{align}
\end{subequations}

\begin{figure*}[t]
\includegraphics[width=17.5cm,height=6.50cm]{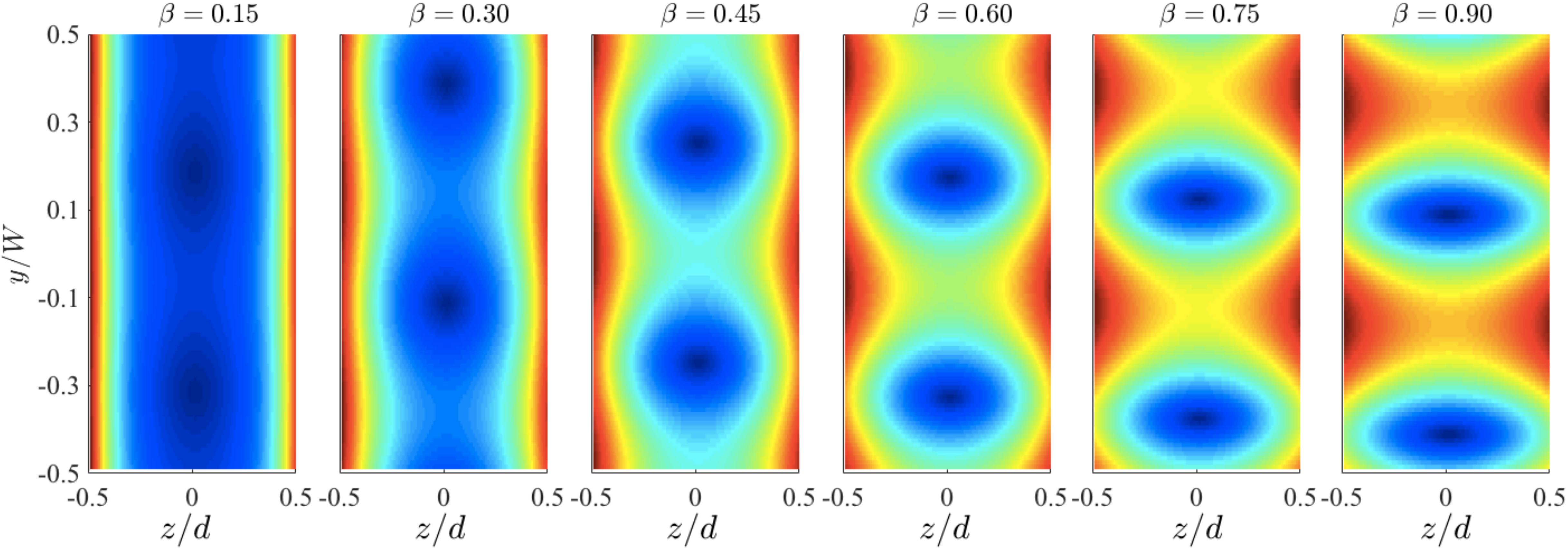}
\caption{\label{fig3} (Color online). 
Spatial map of absolute value of the Cooper pair wavfunction, inside the two-dimensional Josephson junction, normalized by its maximum value (dark blue and dark red are equal to zero and unity, respectively). The coordinates $z$ and $y$ are normalized by the junction length and width, $d$ and $W$, respectively. The superconductor parts are connected to the disordered Weyl semimetal in the $z$ direction along the $y$ axis at $z=\pm d/2$. We set different values to the titling parameter $\beta=0.15, 0.30, 0.45, 0.60, 0.75$, and $0.90$ in the presence of a finite inversion symmetry breaking parameter $\alpha_{z}=0.5$ and $\gamma=0.1$.
 }
\end{figure*}

\begin{equation}\label{crnt2d}
\begin{split}
J_z(y)=e\pi N_0 D^3T\sum_n &\frac{ f_t^2}{\zeta^2\lambda_n} \csc\Big(\frac{\lambda_nD^{-1}}{\beta-\gamma}\Big) \csc\Big(\frac{\lambda_nD^{-1}}{\beta+\gamma}  \Big)\times\\&\left\{ (\beta-\gamma)\sin\Big(\frac{\lambda_nD^{-1}}{\beta+\gamma}  \Big)\sin \Big(\frac{\alpha_z d}{\beta-\gamma}+2{\bm \Phi} y-\varphi\Big) - (\beta+\gamma)\sin\Big(\frac{\lambda_nD^{-1}}{\beta-\gamma}  \Big)\sin \Big(\frac{\alpha_z d}{\beta+\gamma}-2{\bm \Phi} y+\varphi\Big)\right\}.
\end{split}
\end{equation}
%\end{widetext}
\twocolumngrid

To obtain the total charge current flowing through the junction, we integrate the charge current density flowing in the $z$ direction, Eq. (\ref{crnt2d}), over the $y$ direction (perpendicular to the current direction), i.e., $I_c=\int_{-W/2}^{+W/2}J_z(y)dy$, assuming that the junction width is equal to $W$. This integration leads to the standard Fraunhofer diffraction pattern vs the external magnetic flux with a multiplication of Eq. (\ref{crnt1d}). Therefore, the charge current in the presence of an external magnetic field perpendicular to the junction plane has a form of 
\begin{equation}
I_c \propto \Big(\frac{{ \Phi}}{{ \Phi}_0}\Big)^{-1}\sin\Big(\frac{{ \Phi}}{{ \Phi}_0}\Big)[I_1\sin(\varphi_0^++\varphi)+I_2\sin(\varphi_0^-+\varphi)].\end{equation} 
When the inversion symmetry breaking parameter vanishes, the tilting parameter only controls the magnitude of total supercurrent passing through the triplet channel. The presence of inversion symmetry breaking parameter induces more oscillations in the charge current (and thus the Fraunhofer pattern) subject to an external magnetic field when increasing the tilting parameter.

To gain more insight, we plot the spatial map of the absolute value of Cooper pair wave function (i.e. the anomalous Green's function at equal times) within the junction area in Fig. \ref{fig3}. We have set the external flux fixed at ${ \Phi}=2\Phi_0$, $\gamma=0.1$, and the inversion symmetry breaking parameter $\alpha_{z}=0.5$. The rest of parameters are equal to those of Fig. \ref{fig2}. From left to right, we increase the tilting parameter $\beta=0.15, 0.35, 0.45, 0.60, 0.75, 0.90$. The external magnetic field induces two vortices ($n$ vortices for $\Phi=n\Phi_0$) along the junction interface in the $y$ direction at the middle of junction $x=0$ \cite{ma_jap,zu1,ma_odfr}. Increasing $\beta$, the vortices move along the junction width in the $y$ direction. This increase also reforms the vortices so that the point-like cores at $\beta=0.15$ turn to lines with finite sizes expanded in the $z$ direction and the destruction of the pair wavefunction is no longer limited to the junction area and extends into the superconductors. Our further investigates show that varying the strength of inversion symmetry breaking parameter $\alpha_{z}$ only drives the proximity vortices and shifts the locations of vortex cores along the $y$ direction without changing the shape of the vortices' profile whereas varying the tilting parameter when $\alpha_{z}=0$ only changes the shape of vortices the same as what is shown in Fig. \ref{fig3} from left to right, ignoring the location shifts. This is fully consistent with the influences of $\alpha_{z}$ and $\beta$ on the charge current discussed in passing and can be directly inferred from the dependence of charge current density, Eq. (\ref{crnt2d}), on $\beta,\gamma$, and $\alpha_z$. 

\section{Conclusions}\label{sec:cncl}
In summary, we have generalized a quasiclassical model, including the Eilenberger and Usadel equations, for Weyl semimetals with impurities subject to an external magnetic field. As a preliminary step in the application of these generalized techniques to practical systems, we have studied supercurrent flow through a diffusive Weyl semimetal Josephson junction. We have found that the supercurrent can be carried through a triplet channel with a nonzero threshold made of $d\alpha_{z}/(\beta\pm\gamma)$ in which $d$ is the thickness of Weyl semimetal, $\alpha_{z}$ is the strength of inversion breaking parameter, and $\beta,\gamma$ characterize Weyl semimetal. Our results demonstrate that the tilting parameter, $\beta$, can control the induction of current crossovers and the self-biased supercurrent independent of the density of impurities present in the samples. We also consider a two-dimensional Josephson junction and study the effect of $\alpha_{z}$ and $\beta$ on the Cooper pair wavefunction, superconducting vortices, and the response of charge current to an externally applied magnetic field.

\acknowledgments

M.A. is supported by Iran's National Elites Foundation (INEF). M.A. would like to thank A. Zyuzin for useful discussions and a careful reading of the paper. M.A. also thanks G. Sewell for valuable discussions.

\end{document}